\documentclass[twocolumn,prl,superscriptaddress,showpacs]{revtex4}
\usepackage{amsmath,amssymb,graphicx}
\bibstyle{revtex}
\begin{document}

\setlength{\tabcolsep}{1.5mm}
\setcounter{totalnumber}{4}
\setcounter{topnumber}{4}



\title{A multi-blob representation of semi-dilute polymer solutions}

\author{Carlo Pierleoni}
\affiliation{INFM CRS-SOFT and Physics Dept. University of L'Aquila, I-67010 L'Aquila, Italy}

\author{Barbara Capone}
\affiliation{University Chemical Laboratory, Lensfield Road,
Cambridge CB2 1EW, United Kingdom}

\author{Jean-Pierre Hansen}
\affiliation{University Chemical Laboratory, Lensfield Road,
Cambridge CB2 1EW, United Kingdom}


\date{\today}

\begin{abstract}
A coarse-grained multi-blob  description of  polymer solutions is
presented, based on soft, transferable effective interactions between bonded and
non-bonded blobs. The number of blobs is chosen such that the blob
density does not exceed their overlap threshold, allowing polymer concentrations
to be explored deep into the semi-dilute regime. This quantitative multi-blob
description is shown to preserve known scaling laws of polymer solutions and
provides accurate estimates of amplitudes, while leading to orders  of magnitude
increase of simulation efficiency and allowing analytic calculations of
structural and thermodynamic properties.

\end{abstract}

\maketitle

Many conformational, structural and thermodynamic properties  of semi-dilute polymer solutions, both in the bulk and under
confinement, can be qualitatively understood in terms of scaling arguments based on the de Gennes-Pincus ``blob'' picture
\cite{Degennes}. This picture is of value whenever the characteristic length scale of the polymer solution (e.g. the correlation length $\xi$), or of the confinement  is significantly shorter than the radius of gyration $R_g$ of an isolated polymer chain. 
The blob picture suggests a systematic coarse-graining procedure  whereby each polymer chain is divided into a number $n$ of
blobs, each containing the same number of monomers or segments of the initial coil, such that blobs of the same or different
chains do not, on average, overlap. 
In this letter we present and validate a quantitative formulation of such a multi-blob representation, which allows a popular single blob coarse
graining procedure  \cite{Flory, Grosberg, Dautenhahn, Bolhuis, Pelissetto} to be extended to highly concentrated solutions. 

Consider a solution of $N$ self avoiding polymer chains of $L$ segments (each of size $b$) in a volume $V$; the polymer number density is $\rho=N/V$, and if $R_g\sim b L^{\nu}$ (with $\nu\simeq 0.588$ the Flory exponent) is the radius of gyration of an isolated chain, the polymer overlap density is  $\rho^{*}=3/\left(4\pi R_g^3\right)$

In the dilute regime, $\rho^*<\rho$, where polymers do not, on average, overlap, they may be represented by a single blob of radius $ R_g$; the
effective interaction potential $v(r)$ between the centres of mass (CM) of two blobs can be calculated by averaging over monomer conformations for  a
given distance $r$ between their CM's, e.g. by Monte Carlo (MC) simulations of an isolated pair of polymers \cite{Flory,Grosberg,Dautenhahn, Bolhuis,
Pelissetto}. The resulting $v(r)$ depends weakly on polymer length $L$, and in the scaling limit ($L\to\infty$), it is accurately represented by a
Gaussian of width $\sim R_g$ \cite{Pelissetto}:  
\begin{equation}\label{repulsione gaussiana} 
\frac{v(r)}{k_BT} \simeq A \exp \left[-\alpha
(r/R_g)^2\right],  
\end{equation} 
where $A\simeq 1.80$ and $\alpha \simeq 0.80$; the softness of the repulsive interaction, characterised by a modest
free energy penalty of  $\simeq 2 k_B T$ at full overlap ($r=0$) of two polymers, reflects the low average monomer concentration $c\sim L^{1-3\nu}\sim
L^{-0.77}$ inside each coil for long chains. 

In the semi-dilute regime, $\rho > \rho^*$, polymer coils overlap, and this is reflected in a significant density-dependence of the effective interaction \cite{Bolhuis}, which spoils the simplicity of the coarse-graining procedure and introduces complications associated with state-dependent interactions \cite{Louis}. This density-dependence signals the fact that in the semi-dilute regime the relevant length scale is no longer $R_g$, but the shorter correlation length $\xi \sim R_g(\rho/\rho^*)^{-\gamma}$, with $\gamma=\nu/(3\nu-1)\sim 0.77$ \cite{Degennes}.

These shortcomings may be overcome by switching to a multi-blob representation, where each of the $n$ blobs is made up of $l=L/n$ segments. If
$r_g\sim b l^\nu$ is  the blob radius of gyration, the blob overlap concentration is $\rho_b^*=3/\left(4 \pi r_g^3\right)=\rho^* n^{3\nu}$. This means
that the polymer density $\rho=\rho_b /n$ can be increased beyond $\rho^*$, up to $n^{3\nu-1} \rho^*\sim n^{0.77} \rho^*$  before the blobs overlap. 
In other words the more blobs are used to represent one polymer, the deeper one can penetrate into the semi-dilute regime without significant blob overlap. Under those conditions the effective interactions between the CM's of the blobs are expected to be practically independent of blob density $\rho_b$, and may be taken equal to their zero density limit. 

These effective interactions include the pair potential $v(r)$ between non-bonded blobs, and the ``tethering'' potential $\phi(r)$ between adjacent blobs on a given chain. The former is expected to be similar to the Gaussian repulsion in equation \eqref{repulsione gaussiana}, with $R_g$ replaced by $r_g$, i.e. the same as the effective potential between polymers in a single blob representation \cite{Dautenhahn, Bolhuis, Pelissetto}. $\phi(r)$, on the other hand, may be expected to be the superposition of $v(r)$ at short distances $r$, and a harmonic spring at large elongation, similar to the entropic spring of a Gaussian chain \cite{Degennes,Doi}, albeit with a renormalized spring constant:
\begin{equation}\label{tutto il resto}
	\phi(r)=v(r)+\frac{k}{2}(r-r_0)^2 + c.
\end{equation}
These conjectures are borne out by MC simulations of an isolated pair of self-avoiding walk (SAW) polymers on a cubic lattice; each polymer is divided into two sub-chains of equal length $L/2$, and the intra-molecular distribution function $s(r)$ of the CM's of the two blobs, as well as the inter-molecular pair distribution function $g(r)$ of the CM's of blobs on different chains are monitored as functions of the CM-CM distances $r$. 
$\phi(r)$ follows directly from $s(r)\sim \exp (-\phi(r)/k_B T)$, while $v(r)$ is related to $g(r)$ and $s(r)$ via an exact inversion relation \cite{Ladanyi}, which we used previously for simple models of a di-block copolymer \cite{Addison,Pierleoni}. 
The resulting $\phi(r)$ and $v(r)$ are shown in figure 1.
$v(r)$ is indeed of Gaussian form \eqref{repulsione gaussiana}, and is practically indistinguishable from the low density limit of the effective pair potential between the CM's of two polymers of the same length, within the single-blob case \cite{Bolhuis, Pelissetto}. 

\begin{figure}
\includegraphics[width=80mm]{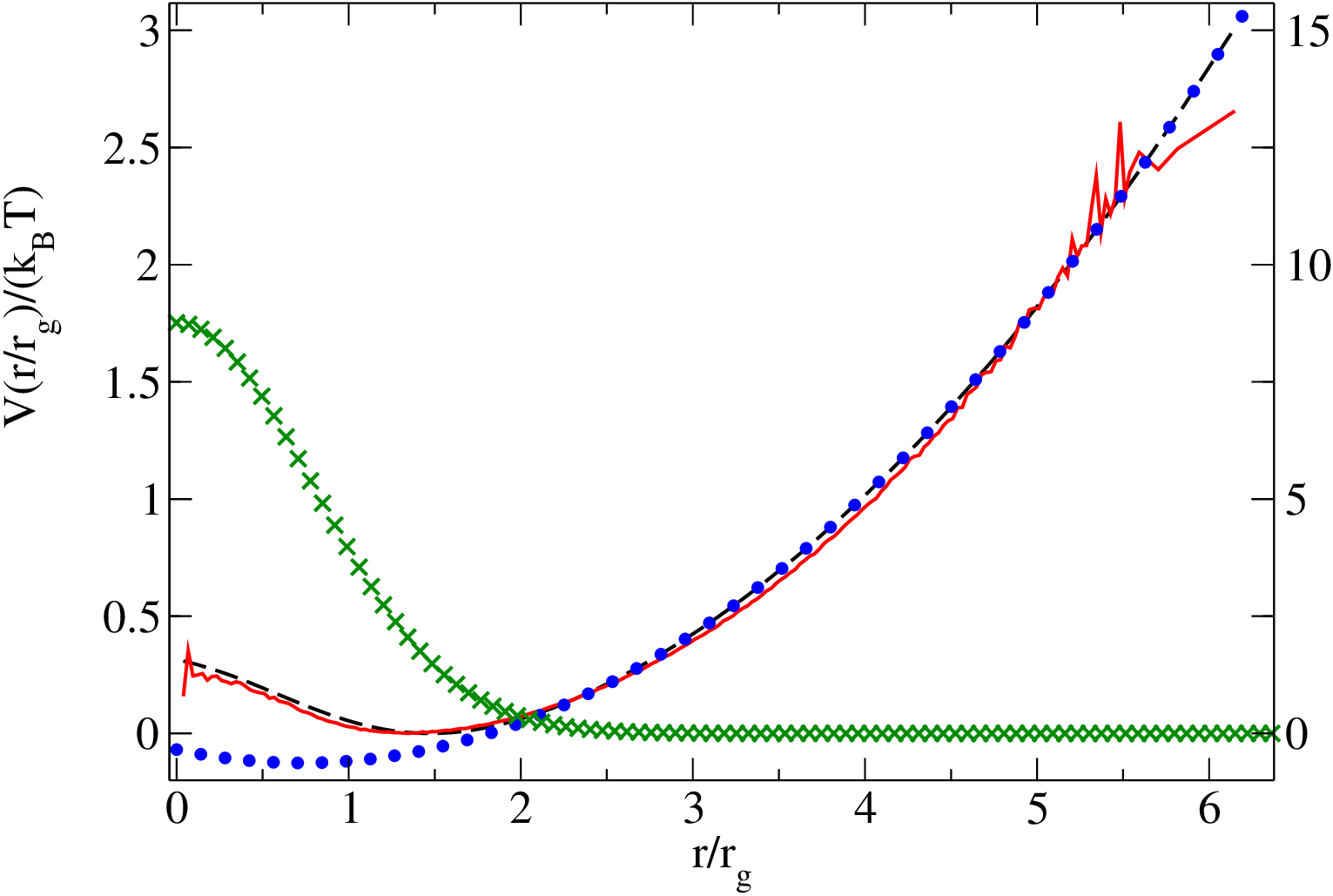}
\caption{\label{figure1} (Color online) Effective potentials $\phi(r)$ between bonded (right hand scale, continuous line)
and v(r) between non-bonded (left-hand scale, crosses) blobs in units of $k_B T$, as computed by MC simulations of an
$L=2000$ SAW polymer, divided into $2$ equal blobs. $v(r)$ is accurately fitted by the Gaussian \eqref{repulsione
gaussiana} with $A=1.752$ and
$\alpha=0.804$. $\phi(r)$ is represented by the form \eqref{tutto il resto} (dashed curve) with the harmonic tethering
potential fitted by  $0.534(r/r_g-0.731)^2-0.631$ (dots).}
\end{figure}

So far no approximation has been made, and the effective potentials in figure \ref{figure1} are ``exact'', for a given
length $L$ of the polymer. As in the single blob representation \cite{Bolhuis, Pelissetto}, the $L$-dependence of $v(r)$ and
$\phi(r)$ is found to be negligible for $L > 1000$. We now make the reasonable transferability assumption that the
interaction potentials in figure \ref{figure1}, as functions of the scaled distance $r/r_g$, remain identical, whatever the
number $n$ of blobs into which a polymer chain is divided, provided the number $l$ of segments in each blob remains
sufficiently large. Note that within this assumption, the potential \eqref{tutto il resto} acts between adjacent (tethered)
blobs only, while $v(r)$ is the same for all pairs of blobs of the same, or different, chains. This defines the multi-blob
model of  homopolymer solutions, and applies as long as $\rho_b < \rho_b^*$. Note that in view of the ``softness'' of the effective interactions the model is particularly well suited for simulation purposes (apart from the obvious speed-up due to the considerable reduction of the number of degrees of freedom from $3N\cdot L$ to $3N\cdot n$) as well as for approximate theories \cite{Schweizer, Louis2}, to which we shall return.

\begin{figure}
\includegraphics[width=80mm]{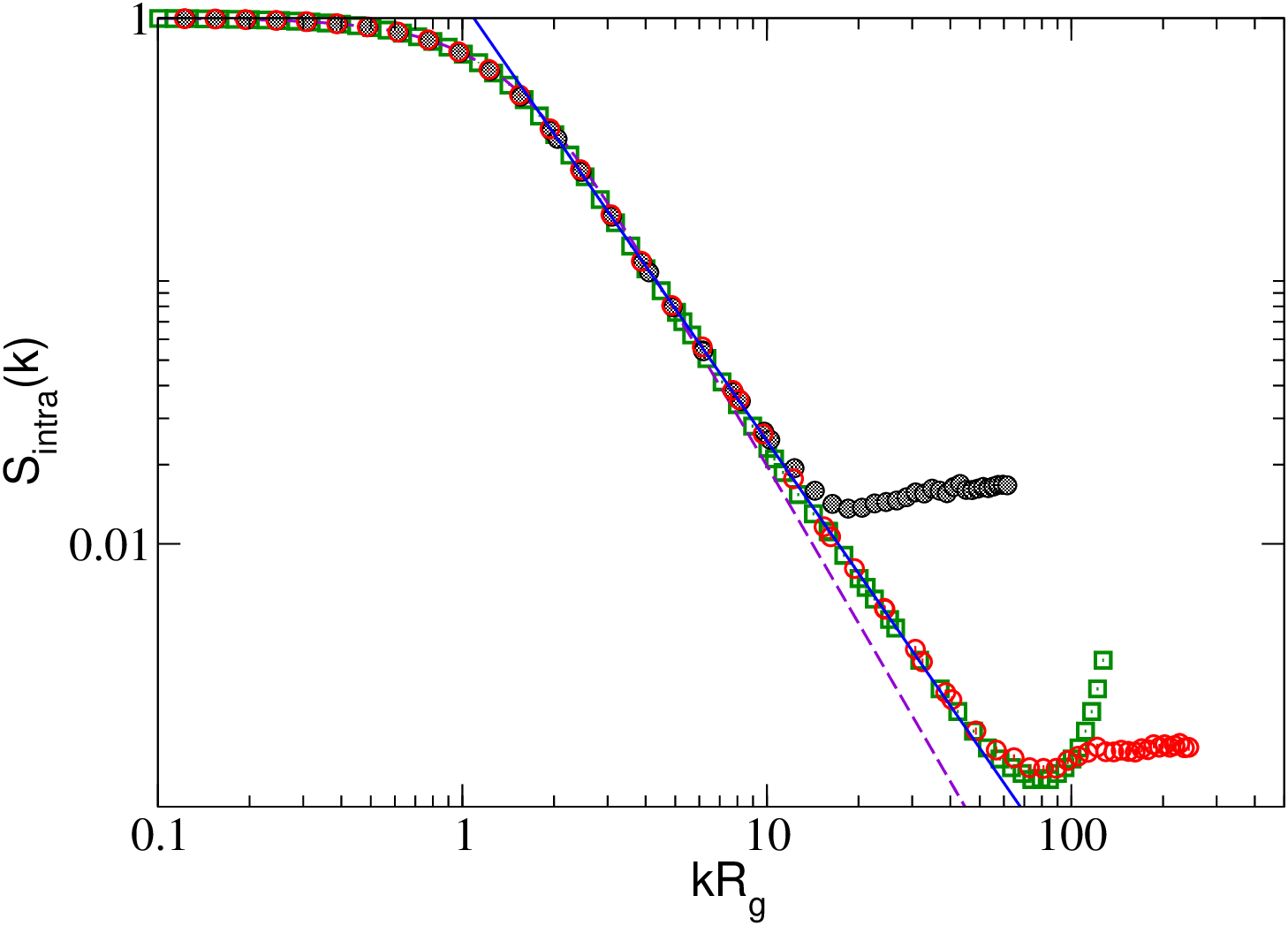}
\caption{\label{secondafigura}(Color online) Intramolecular structure factor $S_\textrm{intra}(k)$ of a single SAW
polymer versus $kR_g$ on a log-log scale. Squares: MC data for an $L=1000$ SAW polymer: closed circles and open circles: MC
data for $n=62$ and $n=602$ blob representations. Full line: scaling regime $1.15(kR_g)^{1/\nu}$, with $\nu\simeq0.5988$:
dashed line: Debye structure factor for a Gaussian chain.}
\end{figure}

We now put the above multi-blob representation to the test, focussing on the structure factor $S(k)$ of a solution of $N$ chains of $n$ blobs each:
\begin{align}\label{fattore di struttura}
\begin{split}
S(k)&=\frac{1}{Nn^2}<\rho_{\vec{k}} \rho_{-\vec{k}}>
\\&=S_\textrm{intra}(k)+\frac{1}{n^2}\sum_{\alpha}\sum_{\beta} s_{\alpha \beta}(k) ,	
\end{split} 
\end{align}
where
\begin{equation}
\rho_{\vec{k}}=\sum_{i=1}^N \sum_{\alpha=1}^n \exp\left(i \vec{k} \cdot \vec{r}_{i\alpha}\right)
\end{equation}
and $\vec{r}_{i\alpha}$ is the position of the centre of blob $\alpha$ on  chain $i$; $s_{\alpha \beta}(k)$ is the partial
intermolecular structure factor for blobs $\alpha$ and $\beta$, while $S_\textrm{intra}(k)$ is the intra-molecular structure factor
 of one polymer. Consider first an isolated polymer of $n$ blobs ($S(k)=S_\textrm{intra}(k)$). Results for the
multi-blob model with $n=62$ and $n=602$ are plotted versus $k \cdot R_g$ on a log-log scale and compared to the structure
factor of a full monomer representation of a SAW with $1000$ monomers in figure \ref{secondafigura}. All the data follow the power law $S(k)\simeq a (k
R_g)^{-1/\nu}$ in the scaling regime $1 < k R_g < R_g/B$, where $B$ is the effective segment length ($r_g$ in the multi-blob representation and $b$ in the full monomer representation), with $\nu \simeq 0.5988$, close to the expected value of the Flory exponent, and $a\simeq 1.15,$ close to the renormalization group prediction \cite{Scaling}. The multi-blob representation does preserve the correct scaling behaviour of the single chain structure factor. The structure factor of the full monomer representation is approximately related to that of a single blob, $s_\textrm{blob}(k)$, and to the structure factor of a chain of blobs, $S_c(k)$ by the factorisation approximation, $S(k)=s_b(k) \cdot S_c(k)$ \cite{Pagonabarraga}.The small $k$ limit of this relation yields an estimate of the radius of gyration $R_g$ of a SAW in terms of those of the chain of blobs ($R_{gc}$) and  of a single blob ($r_g$), namely $R_g^2=R_{gc}^2+r_g^2$.

We next consider $S(k)$ in the semi-dilute regime. Structure factors obtained from MC simulations of $108$  and $500$ chains of $n=5,10,20$ blobs are plotted in figure 3 for $\rho/\rho^*=1.15$ and $9.23$ versus $k R_g$, for a fixed overall length $L$ of the polymer (i.e. a fixed value of $R_g$). 
The overall agreement between the three multi-blob representations is seen to be good for $k r_g < 1$ ($kR_g <n^\nu$) at the lower
density, as one might expect within a consistent coarse-graininig procedure. The significant discrepancy observed at the
higher density illustrates the fact that for $\rho/\rho^*=9.23$, the minimum number $n$ of blobs needed to satisfy the
requirement $\rho_b<\rho_b^*$ is $n=18$. Figure 4 shows the inter-molecular part of the total structure factor \eqref{fattore di struttura}, as calculated with $n=5,10$ or $20$ blobs; the agreement between the various coarse-grained representations is excellent. 
\begin{figure}
\includegraphics[width=80mm]{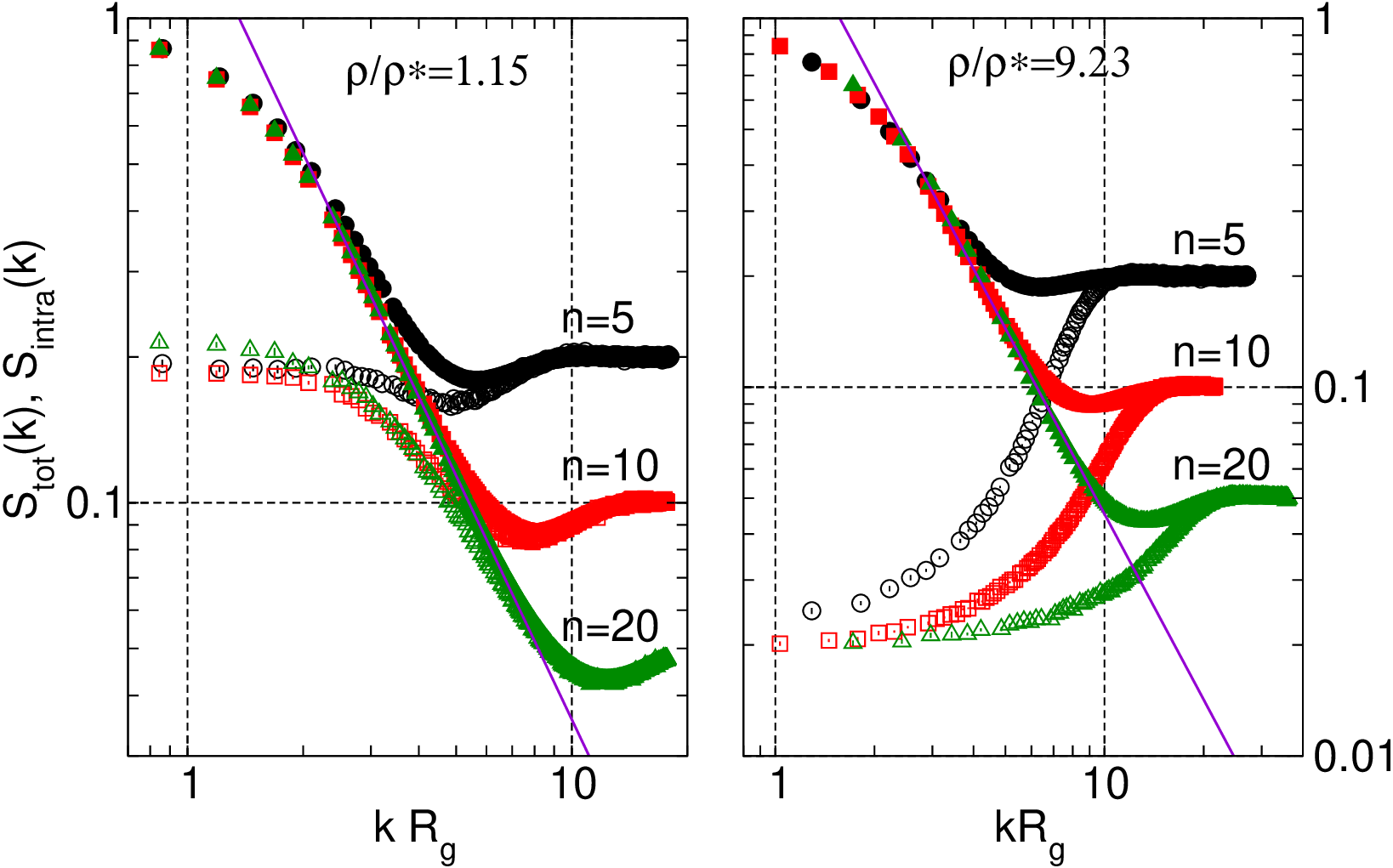}
\caption{(Color online) Total (open symbols) and intramolecular (closed symbols) structure factors versus $kR_g$ from MC simulations of $n=5$
(circles), $n=10$ (squares) and $n=20$ (triangles) blob representations of SAW polymers at $\rho/\rho^*=1.15$ (left frame) and $9.23$ (right frame).
The straight lines correspond to the scaling regime as in fig. \ref{secondafigura} } 
\end{figure}
\begin{figure}
\includegraphics[width=80mm]{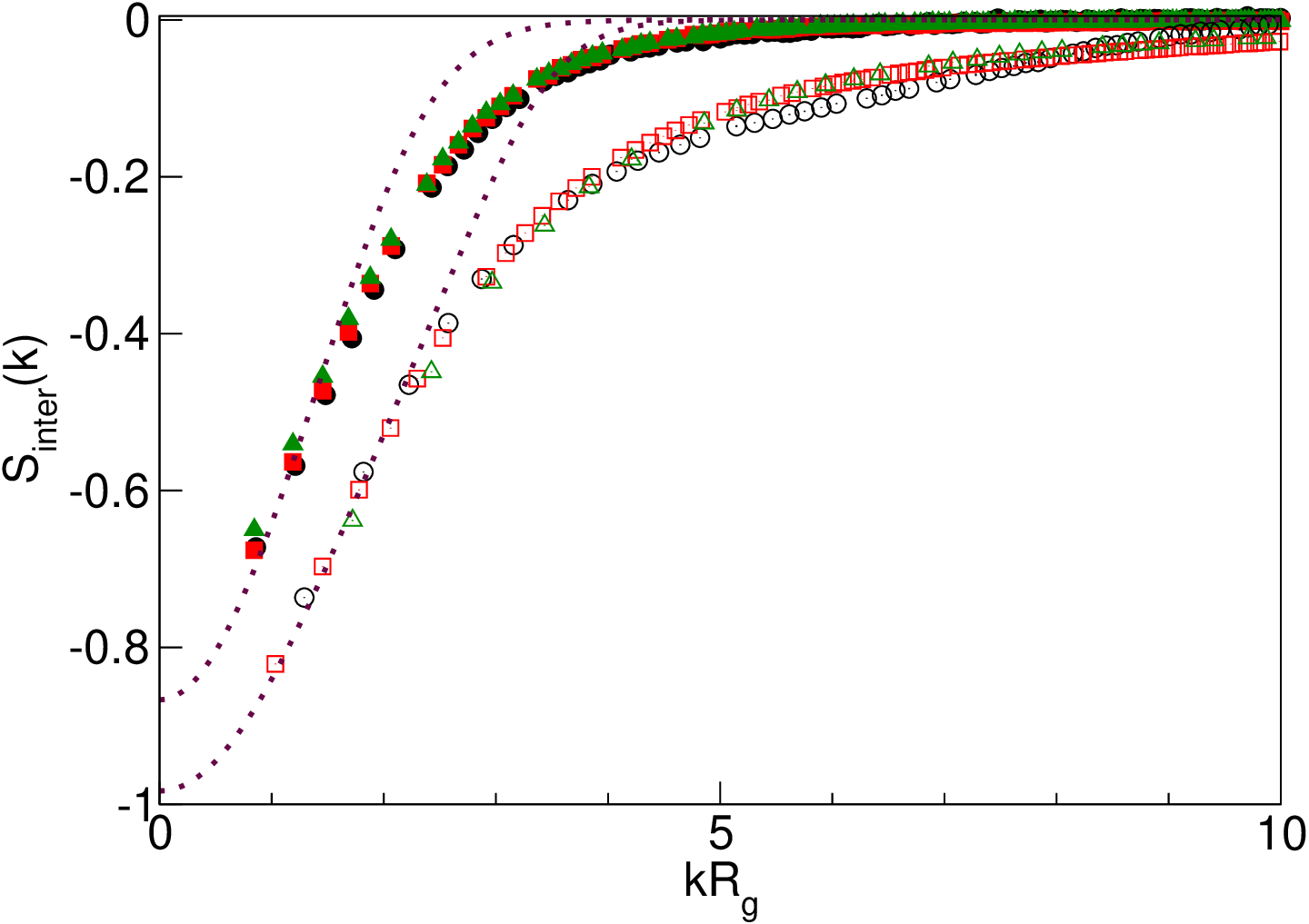}
\caption{ (Color online) Intermolecular structure factor from MC simulation of $m=5$ (circles), $10$ (squares) and $20$ (triangles) blob
representations of SAW polymers at densities $\rho/\rho^*=1.15$ (closed symbols) and $9.23$ (open symbols) versus $k R_g$. The
dashed curves are predictions of the RPA derived from expression \eqref{sei}, using the intramolecular form factor (for
$n=10$). The RPA is seen to be accurate for $k R_g \lesssim 2.5$, but to underestimate short-range correlations ($k R_g \gtrsim 2.5$). }
\end{figure}
The $k\to 0$ limit of the structure factor yields the osmotic compressibility $\chi_T$ via $\rho k_B T \chi_T=S(k \to 0)$. In the same limit all partial intermolecular structure factors become identical so that the compressibility is also given by $\rho k_B T \chi_T = 1 + s_{\alpha \beta}(k \to 0)$. The PRISM approximation assumes all partial structure factors to be identical for all $k$, i.e. $s_{\alpha \beta}(k)=s(k)\,\, \forall\, \alpha \beta$, which is only true in the limit of very long polymers, where end effects may be neglected \cite{Schweizer}. PRISM then relates the total structure factor to the form factor $\hat{\omega}(k)=n\, S_\textrm{intra}(k)$ of the polymer, and to the Ornstein-Zernike direct correlation function $\hat{c}(k)$. For chains of $n$ blobs the relation reads:
\begin{equation}\label{sei}
	S(k)=\frac{\hat{\omega}(k)/n}{1-\rho_b \hat{\omega}(k) \hat{c}(k)}. 
\end{equation}	 
Since the effective interaction between blobs is soft (cf. \eqref{repulsione gaussiana}), the direct correlation function is well approximated by the RPA closure \cite{Louis2} $\hat{c}(k)\simeq - \hat{v}(k)/k_B T$, where hats always refer to Fourier transforms.
Gathering results:
\begin{equation}\label{fattore di forma}
S(k)\simeq \frac{\hat{\omega}(k)/n}{1+\rho_b \hat{\omega}(k)\hat{v}(k)/k_B T}
\end{equation} 
which generalises the classic RPA results for the structure of polymer solutions \cite{Edwards,Daoud,Mueller} to multi-blob chains. The advantage of the result \eqref{sei} is that the RPA on which it relies is accurate because of the weakness of the blob-blob interaction. Using $\hat{\omega}(k=0)=n$ and the Fourier transform of \eqref{repulsione gaussiana} one arrives at the compressibility:
\begin{equation}
	\rho k_B T\chi_T = \frac{1}{1+\rho_b n A \left(\frac{\pi}{\alpha}\right)^{3/2} r_g^3}.
\end{equation}
For given $\rho$ and polymer length $L$, the number $n$ of blobs is chosen such that $\rho_b=n\rho=\rho_b^*$, i.e. $n=\left(\rho/\rho^*\right)^{1/(3\nu-1)}$.
Hence: 
\begin{equation}\label{sette}
\rho k_B T \chi_T = \frac {1} {1 + \zeta \left(\rho/\rho^*\right)^{1/(3 \nu -1)}},
\end{equation} 
where $\zeta =3 A \left(\pi/\alpha\right)^{3/2} / 4\pi \simeq 3.23$.
For $\rho>>\rho^*$ \eqref{sette} leads back to the des Cloizeaux scaling for the osmotic pressure $\Pi \sim (\rho/\rho^*)^{3 \nu / (3 \nu -1)}\sim\left(\rho/\rho^*\right)^{2.3}$ \cite{Degennes}.
Numerical values obtained for the compressibility from eq \eqref{sette} agree with MC data (as shown in figure 3) within a few percent, except at the
highest density where a discrepancy of $20\%$ is probably due to large statistical uncertainties in the MC data. Another consequence of the weakness of the effective interaction between blobs is that solutions of multi-blob chains lend themselves readily to thermodynamic perturbation theory \cite{Hansen}. 
Choosing a system of non-interacting Gaussian coils of length $n$ and spring constant $k=3k_B T/B_0^2$ \cite{Doi} as reference system, the free energy $F$ of a solution of an equal number of $n$-blob chains satisfies the Gibbs-Bogoliubov inequality \cite{Hansen}
\begin{equation}\label{otto}
	F\leq F_0+ \left<V_n-V_n^{(0)}\right>_0
\end{equation}
where $F_0$ is the free energy of the reference system, $V_n$ and $V_n^{(0)}$ are the total energies of the interacting n-blob system and non-interacting reference system, and the average is taken with Gaussian statistics. 
The two terms on the right hand side are readily calculated analytically as functions of the bond length $B_0$ of the Gaussian chains. The right hand side of \eqref{otto} is then minimised with respect to $B_0$ to provide the best variational estimate of $F$. 
The optimum $B_0$ leads directly to an estimate of the radius of gyration of interacting polymers in a semi-dilute solution, namely $R_g=n^{0.5} B_0/\sqrt{6}=n^{0.5} b_0 r_g/\sqrt{6}$, where $r_g\simeq (L/n)^\nu$ is the radius of gyration of one blob and $b_0=(B_0/r_g)$ is determined by the free energy minimisation; $b_0$ increases with $n$ and saturates at a value $b_0\simeq 2.24$ for $n > 100 $. For a given polymer density $\rho$ and length $L$, $n$ is once more determined by the requirement that the blob density $\rho_b=n \rho$ equals its overlap value $\rho^*_b$; this leads to the desired result: 
\begin{equation}\label{nove}
	R_g=c b L^{\nu} \left(\frac{\rho}{\rho^*}\right)^{\frac{1-2\nu}{2 (3\nu -1)}}
\end{equation}
where $c\simeq 0.40 $ for sufficiently large $n$. Eq. \eqref{nove} shows the correct slow decrease of $R_g$ with density $(R_g \simeq (\rho/\rho^*)^{-0.115})$ predicted by scaling theory \cite{Degennes}. Note that in the melt limit $n\to L$, $R_g$ scales like $L^{1/2}$, as expected. 

The present multi-blob coarse-grained description provides a  fresh look at semi-dilute polymer solutions and allows earlier work on dilute solutions \cite{Degennes, Flory, Grosberg, Dautenhahn, Bolhuis, Pelissetto} to be extended to concentrated solutions of long chains, using transferable, state-independent pair interactions between blobs. The softness of the latter, and reduction of the number of degrees of freedom, lead, on one hand, to a speed-up of simulations by orders of magnitude, thus allowing the simulation of large numbers of high molecular mass polymers, while, on the other hand, opening up the systematic use of standard methods of statistical theories of fluids. We have shown that the coarse-graining procedure preserves standard scaling laws of semi-dilute polymer solutions, while providing reliable estimates of amplitudes (i.e. numerical pre-factors). We are presently extending the  multi-blob methodology to more general homopolymer models involving effective attractions between adjacent monomers (and hence solvent quality or temperature effects), to brushes of grafted polymers \cite{Coluzza}, to colloid-polymer mixtures and to the self-assembly of block copolymer solutions, going beyond the minimal two-blob representation of diblock copolymers used in references\cite{Addison, Pierleoni}.
\section*{Acknowledgments}
We thank Peter Bolhuis and Vincent Krakoviack for valuable comments. BC is supported by an EC Marie-Curie studentship within the Eurosim Network. CP is grateful to the Royal Society for a short visit grant.

\bibliography{biblio}

\begin{thebibliography}{20}
\expandafter\ifx\csname natexlab\endcsname\relax\def\natexlab#1{#1}\fi
\expandafter\ifx\csname bibnamefont\endcsname\relax
  \def\bibnamefont#1{#1}\fi
\expandafter\ifx\csname bibfnamefont\endcsname\relax
  \def\bibfnamefont#1{#1}\fi
\expandafter\ifx\csname citenamefont\endcsname\relax
  \def\citenamefont#1{#1}\fi
\expandafter\ifx\csname url\endcsname\relax
  \def\url#1{\texttt{#1}}\fi
\expandafter\ifx\csname urlprefix\endcsname\relax\def\urlprefix{URL }\fi
\providecommand{\bibinfo}[2]{#2}
\providecommand{\eprint}[2][]{\url{#2}}

\bibitem[{\citenamefont{de~Gennes}(1979)}]{Degennes}
\bibinfo{author}{\bibfnamefont{P.~G.} \bibnamefont{de~Gennes}},
  \emph{\bibinfo{title}{Scaling Concepts in Polymer Physics}}
  (\bibinfo{publisher}{Cornell University Press,(Ithaca)},
  \bibinfo{year}{1979}).

\bibitem[{\citenamefont{Flory and Krigbaum}(1950)}]{Flory}
\bibinfo{author}{\bibfnamefont{P.~J.} \bibnamefont{Flory}} \bibnamefont{and}
  \bibinfo{author}{\bibfnamefont{W.~R.} \bibnamefont{Krigbaum}},
  \bibinfo{journal}{J. Chem. Phys} \textbf{\bibinfo{volume}{18}},
  \bibinfo{pages}{1086} (\bibinfo{year}{1950}).

\bibitem[{\citenamefont{Grosberg et~al.}(1982)\citenamefont{Grosberg, Khalatur,
  and Khoklov}}]{Grosberg}
\bibinfo{author}{\bibfnamefont{A.~V.} \bibnamefont{Grosberg}},
  \bibinfo{author}{\bibfnamefont{P.~G.} \bibnamefont{Khalatur}},
  \bibnamefont{and} \bibinfo{author}{\bibfnamefont{A.~R.}
  \bibnamefont{Khoklov}}, \bibinfo{journal}{Makromol. Chem. Rapid Commun}
  \textbf{\bibinfo{volume}{3}}, \bibinfo{pages}{709} (\bibinfo{year}{1982}).

\bibitem[{\citenamefont{Dautenhahn and Hall}(1994)}]{Dautenhahn}
\bibinfo{author}{\bibfnamefont{J.}~\bibnamefont{Dautenhahn}} \bibnamefont{and}
  \bibinfo{author}{\bibfnamefont{C.~K.} \bibnamefont{Hall}},
  \bibinfo{journal}{Macromol.} \textbf{\bibinfo{volume}{27}},
  \bibinfo{pages}{5399} (\bibinfo{year}{1994}).

\bibitem[{\citenamefont{Bolhuis et~al.}(2001)\citenamefont{Bolhuis, Louis,
  Hansen, and Meyer}}]{Bolhuis}
\bibinfo{author}{\bibfnamefont{P.~G.} \bibnamefont{Bolhuis}},
  \bibinfo{author}{\bibfnamefont{A.~A.} \bibnamefont{Louis}},
  \bibinfo{author}{\bibfnamefont{J.~P.} \bibnamefont{Hansen}},
  \bibnamefont{and} \bibinfo{author}{\bibfnamefont{E.~J.} \bibnamefont{Meyer}},
  \bibinfo{journal}{J. Chem. Phys.} \textbf{\bibinfo{volume}{114}},
  \bibinfo{pages}{4296} (\bibinfo{year}{2001}).

\bibitem[{\citenamefont{Pelissetto and Hansen}(2005)}]{Pelissetto}
\bibinfo{author}{\bibfnamefont{A.}~\bibnamefont{Pelissetto}} \bibnamefont{and}
  \bibinfo{author}{\bibfnamefont{J.~P.} \bibnamefont{Hansen}},
  \bibinfo{journal}{J. Chem. Phys.} \textbf{\bibinfo{volume}{112}},
  \bibinfo{pages}{134904} (\bibinfo{year}{2005}).

\bibitem[{\citenamefont{Louis}(2002)}]{Louis}
\bibinfo{author}{\bibfnamefont{A.~A.} \bibnamefont{Louis}},
  \bibinfo{journal}{J. Phys. Cond. Mat.} \textbf{\bibinfo{volume}{14}},
  \bibinfo{pages}{9187} (\bibinfo{year}{2002}).

\bibitem[{\citenamefont{Doi}(1996)}]{Doi}
\bibinfo{author}{\bibfnamefont{M.}~\bibnamefont{Doi}},
  \emph{\bibinfo{title}{Introduction to Polymer Physics}}
  (\bibinfo{publisher}{Clarendon Press,(Oxford)}, \bibinfo{year}{1996}).

\bibitem[{\citenamefont{Ladanyi and Chandler}(1975)}]{Ladanyi}
\bibinfo{author}{\bibfnamefont{B.~M.} \bibnamefont{Ladanyi}} \bibnamefont{and}
  \bibinfo{author}{\bibfnamefont{D.}~\bibnamefont{Chandler}},
  \bibinfo{journal}{J. Chem. Phys.} \textbf{\bibinfo{volume}{62}},
  \bibinfo{pages}{4308} (\bibinfo{year}{1975}).

\bibitem[{\citenamefont{Addison et~al.}(2005)\citenamefont{Addison, Hansen,
  Krakoviack, and Louis}}]{Addison}
\bibinfo{author}{\bibfnamefont{C.~I.} \bibnamefont{Addison}},
  \bibinfo{author}{\bibfnamefont{J.~P.} \bibnamefont{Hansen}},
  \bibinfo{author}{\bibfnamefont{V.}~\bibnamefont{Krakoviack}},
  \bibnamefont{and} \bibinfo{author}{\bibfnamefont{A.~A.} \bibnamefont{Louis}},
  \bibinfo{journal}{Molec. Phys.} \textbf{\bibinfo{volume}{103}},
  \bibinfo{pages}{3045} (\bibinfo{year}{2005}).

\bibitem[{\citenamefont{Pierleoni et~al.}(2006)\citenamefont{Pierleoni,
  Addison, Hansen, and Krakoviack}}]{Pierleoni}
\bibinfo{author}{\bibfnamefont{C.}~\bibnamefont{Pierleoni}},
  \bibinfo{author}{\bibfnamefont{C.~I.} \bibnamefont{Addison}},
  \bibinfo{author}{\bibfnamefont{J.~P.} \bibnamefont{Hansen}},
  \bibnamefont{and}
  \bibinfo{author}{\bibfnamefont{V.}~\bibnamefont{Krakoviack}},
  \bibinfo{journal}{Phys. Rev. Lett.} \textbf{\bibinfo{volume}{96}},
  \bibinfo{pages}{128302} (\bibinfo{year}{2006}).

\bibitem[{\citenamefont{Shweizer and Curro}(1997)}]{Schweizer}
\bibinfo{author}{\bibfnamefont{K.~S.} \bibnamefont{Shweizer}} \bibnamefont{and}
  \bibinfo{author}{\bibfnamefont{J.~G.} \bibnamefont{Curro}},
  \bibinfo{journal}{Adv. Chem. Phys.} \textbf{\bibinfo{volume}{98}},
  \bibinfo{pages}{1} (\bibinfo{year}{1997}).

\bibitem[{\citenamefont{Louis et~al.}(2000)\citenamefont{Louis, Bolhuis, and
  Hansen}}]{Louis2}
\bibinfo{author}{\bibfnamefont{A.~A.} \bibnamefont{Louis}},
  \bibinfo{author}{\bibfnamefont{P.~G.} \bibnamefont{Bolhuis}},
  \bibnamefont{and} \bibinfo{author}{\bibfnamefont{J.~P.}
  \bibnamefont{Hansen}}, \bibinfo{journal}{Phys. Rev. E}
  \textbf{\bibinfo{volume}{62}}, \bibinfo{pages}{7961} (\bibinfo{year}{2000}).

\bibitem[{\citenamefont{des Cloizeaux and Jannink}(1990)}]{Scaling}
\bibinfo{author}{\bibfnamefont{J.}~\bibnamefont{des Cloizeaux}}
  \bibnamefont{and} \bibinfo{author}{\bibfnamefont{G.}~\bibnamefont{Jannink}},
  \emph{\bibinfo{title}{Polymers in Solution, their Modelling and Structure}}
  (\bibinfo{publisher}{Oxford University Press}, \bibinfo{year}{1990}).

\bibitem[{\citenamefont{Pagonabarraga and Cates}(2001)}]{Pagonabarraga}
\bibinfo{author}{\bibfnamefont{I.}~\bibnamefont{Pagonabarraga}}
  \bibnamefont{and} \bibinfo{author}{\bibfnamefont{M.~E.} \bibnamefont{Cates}},
  \bibinfo{journal}{Europhys. Lett.} \textbf{\bibinfo{volume}{55}},
  \bibinfo{pages}{348} (\bibinfo{year}{2001}).

\bibitem[{\citenamefont{Edwards}(1966)}]{Edwards}
\bibinfo{author}{\bibfnamefont{S.~F.} \bibnamefont{Edwards}},
  \bibinfo{journal}{Proc. Phys. Soc. London} \textbf{\bibinfo{volume}{88}},
  \bibinfo{pages}{265} (\bibinfo{year}{1966}).

\bibitem[{\citenamefont{Daoud and et~al}(1975)}]{Daoud}
\bibinfo{author}{\bibfnamefont{M.}~\bibnamefont{Daoud}} \bibnamefont{and}
  \bibinfo{author}{\bibnamefont{et~al}}, \bibinfo{journal}{Macromolecules}
  \textbf{\bibinfo{volume}{8}}, \bibinfo{pages}{804} (\bibinfo{year}{1975}).

\bibitem[{\citenamefont{Mueller et~al.}(2000)\citenamefont{Mueller, Binder, and
  Schaefer}}]{Mueller}
\bibinfo{author}{\bibfnamefont{M.}~\bibnamefont{Mueller}},
  \bibinfo{author}{\bibfnamefont{K.}~\bibnamefont{Binder}}, \bibnamefont{and}
  \bibinfo{author}{\bibfnamefont{L.}~\bibnamefont{Schaefer}},
  \bibinfo{journal}{Macromolecules} \textbf{\bibinfo{volume}{33}},
  \bibinfo{pages}{4568} (\bibinfo{year}{2000}).

\bibitem[{\citenamefont{Hansen and Mc~Donald}(2006)}]{Hansen}
\bibinfo{author}{\bibfnamefont{J.~P.} \bibnamefont{Hansen}} \bibnamefont{and}
  \bibinfo{author}{\bibfnamefont{I.~R.} \bibnamefont{Mc~Donald}},
  \emph{\bibinfo{title}{Theory of Simple Liquids, 3rd edition}}
  (\bibinfo{publisher}{Academic Press,(Amsterdam)}, \bibinfo{year}{2006}).

\bibitem[{\citenamefont{Coluzza and Hansen}(2007)}]{Coluzza}
\bibinfo{author}{\bibfnamefont{I.}~\bibnamefont{Coluzza}} \bibnamefont{and}
  \bibinfo{author}{\bibfnamefont{J.~P.} \bibnamefont{Hansen}},
  \bibinfo{journal}{to be published}  (\bibinfo{year}{2007}).

\end{thebibliography}

\end{document}